\documentclass[prb,twocolumn,showpacs]{revtex4}
\usepackage{epsfig,times,amsmath,amssymb,color}
\begin{document}

\title{Exotic paired phases in ladders with spin-dependent hopping}
\author{Adrian E. Feiguin}
\affiliation{Department of Physics and Astronomy, University of Wyoming, Laramie, WY 82071, USA}

\author{Matthew P. A. Fisher}
\affiliation{Department of Physics, California Institute of Technology, MC 114-36, Pasadena, CA 91125, USA}

\date{\today}

\begin{abstract}
Fermions in two-dimensions (2D) when subject to anisotropic spin-dependent hopping  can potentially give rise to 
unusual paired states in {\it unpolarized} mixtures that can behave as non-Fermi liquids.
One possibility is a fully paired state with a gap for fermion excitations in which the
Cooper pairs remain uncondensed.  Such a ``Cooper-pair Bose-metal" phase 
would be expected to have a singular Bose-surface in momentum space.
As demonstrated in the context of 2D bosons hopping with a frustrating ring-exchange interaction,
an analogous Bose-metal phase has a set of quasi-1D descendent states when put on a ladder geometry.   Here we present a density matrix renormalization group (DMRG) study of the attractive Hubbard model with spin-dependent hopping on a two-leg ladder geometry. In our setup, one spin species moves preferentially along the leg direction, while the other does so along the rung direction. 
We find compelling evidence for the existence of a novel Cooper-pair Bose-metal phase
in a region of the phase diagram at intermediate coupling.  We further explore the phase diagram of this model as a function of hopping anisotropy, density, and interaction strength, finding a conventional superfluid phase, as well as a phase of paired Cooper pairs with d-wave symmetry, similar to the one found in models of hard-core bosons with ring-exchange. We argue that simulating this model with cold Fermi gases on spin dependent optical lattices is a promising direction for realizing exotic quantum states.
\end{abstract}
\maketitle

\section{Introduction}

The quest for exotic phases of matter of quantum origin is one of the most exciting topics in modern condensed matter physics. Very recently, the extraordinary progress in experiments with cold atomic gases has motivated efforts toward realizing artificial Hamiltonians in a lab, under controlled experimental conditions\cite{Z-K-review}. These Hamiltonians --close realizations of paradigmatic models such as the Bose-Hubbard model\cite{Fisher89,Greiner,Greiner02}-- could, in turn, display very rich physics that may, or may not be present in actual condensed matter systems. The ability to tune the interactions and hopping parameters, even complex ring-exchange terms\cite{Buchler05}, or artificial vector potentials\cite{Spielman09}, allows for an unprecedented freedom to explore new uncharted territory.

A very interesting avenue to explore is the realization of non-Fermi liquids. One possibility is a state formed by bosonic Cooper pairs that cannot condense due to the presence of frustration. The bosons would then behave as a ``normal" fluid, instead of a superfluid. Realizing and understanding such a state could help to shed light on fundamental aspects of the physics of pairing. Recently, in a series of papers \cite{Motrunich07,Sheng08,Sheng09}, one of the authors and collaborators proposed a ``d-wave correlated Bose metal" (DBM) state in terms of bosons. This itinerant uncondensed state is constructed by writing a boson in terms of fermionic partons with anisotropic Fermi surfaces. In this paper we propose realizing such a DBM state in an optical lattice, but using real fermions as the constituents of the fluid. 

\begin{figure}
\epsfig {file=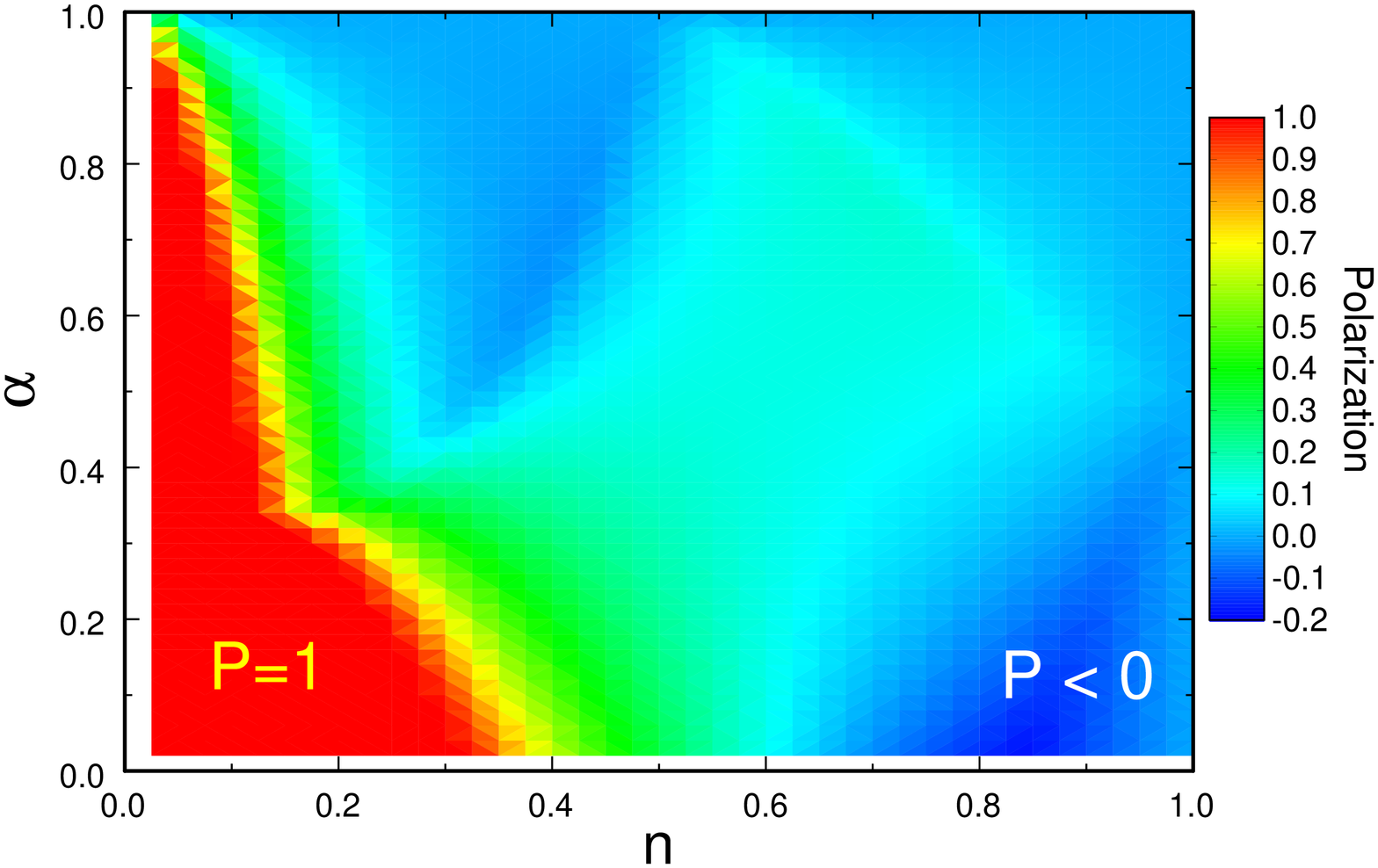,height=80mm,angle=-90}
\epsfig {file=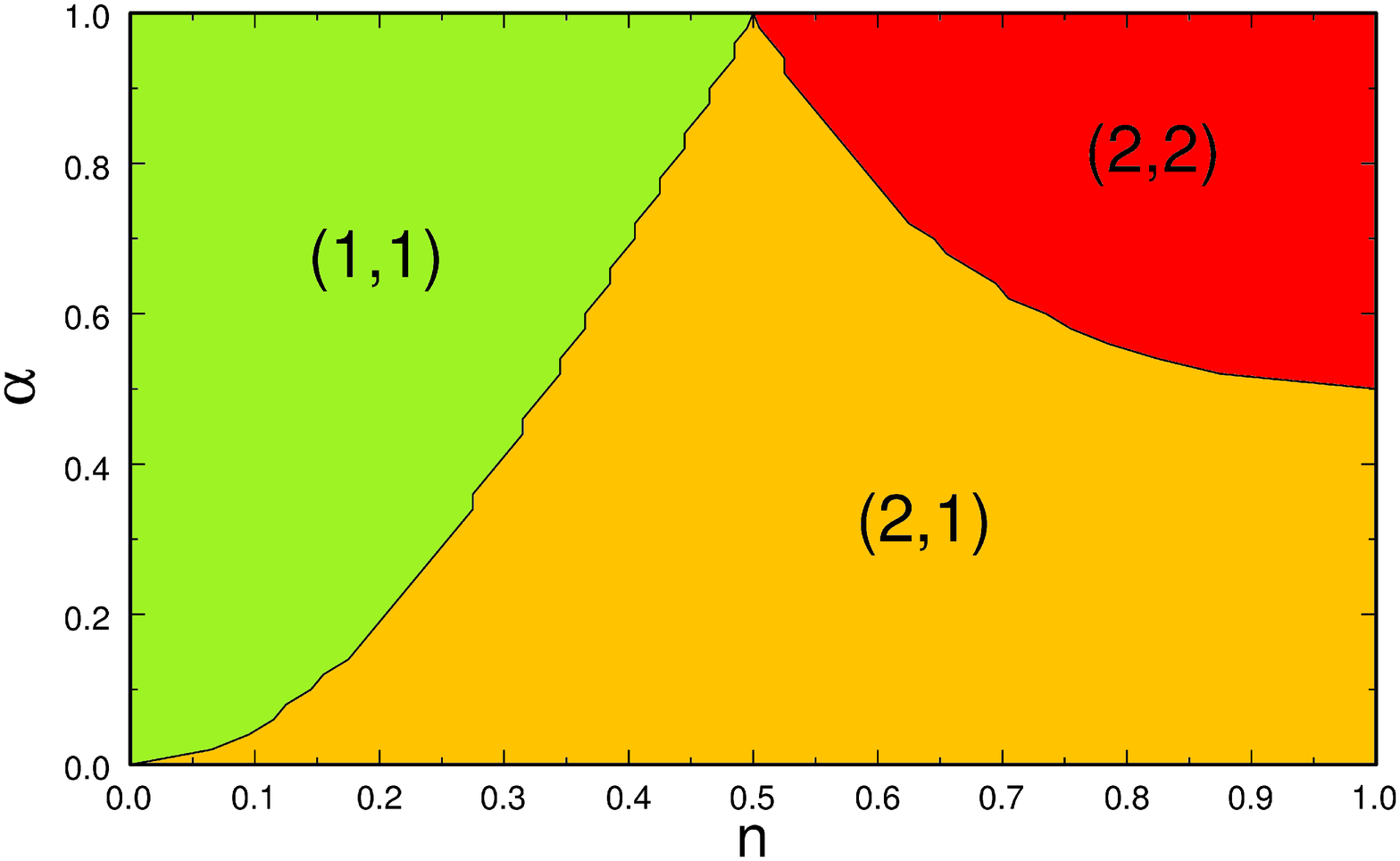,height=80mm,angle=-90}
\caption{
Phase diagram for non-interacting ($U=0$) fermions with anisotropic hoppings on a ladder geometry, as a function of anisotropy $\alpha$ and density $n$. The upper panel shows the polarization of the ground state, while the lower panel shows the different ``phases'' in the unpolarized situation, in terms of the band fillings. 
} \label{bands}
\end{figure}
In a previous paper \cite{Feiguin09}, we suggested a setup to access unconventional paired states in ultracold fermionic systems. Consider an experiment with two fermionic hyperfine states ($\uparrow$,$\downarrow$), that move on a square lattice. We use spin-dependent optical lattices to tune the hoppings such that one species moves preferentially  along the $x$ direction, while the other moves preferentially along the $y$ direction. We consider for simplicity a situation in which the respective Fermi surfaces are rotated by 90 degrees, but the main ingredient is to have mismatched Fermi surfaces.
When one turns on a short range s-wave attractive interaction between the fermions, Cooper pairs can form.
In contrast to the spin imbalanced mixtures that can lead to Fulde-Ferrell-Larkin-Ovchinnikov (FFLO) superconductivity \cite{FF,LO,Leo,Leo2}, in our case we focus on an {\it unpolarized} gas.
But due to the mismatch in Fermi momenta, different pairing solutions are possible.
With a strong attractive interaction all of the fermions can pair and condense into a 
conventional superfluid with zero center of mass momentum.
Alternatively we can realize a gapless state, similar to the Sarma or breached-paired (BP) state for polarized mixtures, with coexistence of pairs and unpaired fermions \cite{Sarma,breached1,breached2}.   

In Ref.\onlinecite{Feiguin09} we explored the BCS mean field phase diagram for such a problem, and found that the gapless superfluid is a stable solution in a wide region of parameter space. 
But an even more exotic possibility would
be a state in which all of the fermions are paired into Cooper pairs, but the Cooper pairs
remain in an uncondensed non-superfluid phase.  This 
``Cooper-pair Bose metal" (CPBM) is not accessible in a BCS mean-field treatment, or any other weak coupling approach.  
Even though there are no a priori arguments to prevent a CPBM state from occurring, 
accessing such a phase would necessarily require a strong coupling treatment.
To this end, in this work we use the density matrix renormalization group method\cite{dmrg} to explore the phase diagram of the same model on a two-leg ladder geometry, as a function of density, anisotropy, and interaction strength.


\subsection{Background}

In our previous work \cite{Feiguin09}, we studied the effects of an attractive interaction in a Fermi mixture with anisotropic Fermi surfaces. In particular, we focused on the situation where the Fermi surfaces of the two spin states
are rotated by $90$ degrees with respect to one another. The resulting model can be described by a generalized Hubbard Hamiltonian with {\it spin-dependent} near neighbor hopping $t_{x,\sigma}$, $t_{y,\sigma}$. By simply taking $t_{y\downarrow}=t_{x\uparrow}=t$, $t_{x\downarrow}=t_{y\uparrow}=\alpha t$ we obtain single particle dispersions;
\begin{eqnarray}
\epsilon_\uparrow(k_x,k_y)=-2t\cos{(k_x)}-2\alpha t\cos{(k_y)},\\
\epsilon_\downarrow(k_x,k_y)=-2\alpha t \cos{(k_x)}-2t\cos{(k_y)}.
\label{omega}
\end{eqnarray}
The parameter $\alpha$, which we take between zero and one, determines the
eccentricity of the two Fermi surfaces.

We assumed that the particles interact through a short-range s-wave
potential, that we represent using the attractive Hubbard model:
\begin{eqnarray}
H =  & \sum\limits_{k,{\sigma}} \epsilon_\sigma(k) c^\dagger_{k,\sigma}
c_{k,\sigma} +
U \sum\limits_{i} n_{i,\uparrow}n_{i,\downarrow}
~, \label{one}
\end{eqnarray}
where $c^\dagger_{k,\sigma}$ creates a fermion with spin
$\sigma=\,\uparrow ,\downarrow$ at momentum $k$,
$n_{i\sigma}=c^\dagger_{i\sigma}c_{i \sigma}$ is the local
on-site density, and $U$ is the
interaction strength that we will take to be negative
(attractive).   

For a very strong attractive interaction, $|U|>> t$, a state with zero
momentum pairing and a fully gapped Fermi surface is expected.
For smaller $U$ a zero-momentum paired state with
gapless single fermion excitations, analogous
to the Sarma or BP phase in mass imbalanced mixtures, could occur.
Alternatively, pairing could occur across the
two mismatched Fermi surfaces, leading to a superfluid state with finite
center-of-mass momentum. The resulting order parameter would be the same as
in the FFLO state with a spatially modulated
condensate at some non-zero wave vector $Q$.

A more exotic possibly that we suggested is a Cooper-pair Bose-metal, a state in which the fermions are fully paired with a fermion gap, but the Cooper pairs have nevertheless not condensed.
Rather, the Cooper pairs form a non-superfluid ``Bose-metal".  As explained in Ref. \onlinecite{Motrunich07},
our motivation for considering the CPBM phase was based upon a mapping to an effective
boson model in the $|U| >> t $ limit.  In addition to the usual boson hopping term with strength
$J \sim \alpha t^2/|U|$, one obtains a 4-site ring exchange term with strength $K \sim t^4/|U|^3$;
\begin{equation}
H_{\mathrm{ring}} = K \sum_{\mathrm{plaquettes}} b^\dagger_1 b_2 b^\dagger_3 b_4 + {\mathrm H.c.} ,
\end{equation}
with $i=1,2,3,4$ labeling sites taken clockwise around a square plaquette.  
Here, $b_i = c_{i \uparrow} c_{i \downarrow}$.
Importantly, while $J \sim \alpha$ vanishes with the anisotropy parameter,
the ring term is independent of $\alpha$ for $\alpha \rightarrow 0$.   Thus, with large Fermi surface anisotropy one expects the ring term to become more important.
In Ref. \onlinecite{Motrunich07} and \onlinecite{Sheng08} it was established that the presence of such a ring term can lead to the existence
of an exotic unpaired Bose-metal phase, referred to as a d-wave Bose metal.   Extensive numerics were done on the two-leg ladder to establish this, which were bolstered by a parton construction
wherein the boson was expressed as a product of two fermionic partons.   

Here we are interested in using real fermions as the constituents of the fluid,
and will be interested whether they can pair and form an analogous Bose metal,
but made of Cooper pairs.  To be specific, in this work we study numerically a version of this model on a two-leg ladder geometry using the DMRG method, which is an unbiased technique that allows one to study large quasi one-dimensional systems with extraordinary accuracy\cite{dmrg}.   Our main findings is that in a range of intermediate coupling
with $|U|/t \sim 4$  we find strong evidence for the existence of the 2-leg ladder
descendent of the Cooper-pair Bose-metal.

\section{Anisotropic hoppings on a ladder geometry}

Throughout we will study the Hubbard Hamiltonian
\begin{eqnarray}
H = & - & \sum_{i,\lambda,\sigma}t_{x,\sigma}\left(c^\dagger_{i,\lambda,\sigma} c^{\phantom{\dagger}}_{i+1,\lambda,\sigma}+{\mathrm H.c.}\right)\nonumber \\
 & - & \sum_{i,\sigma}t_{y,\sigma}\left(c^\dagger_{i,1,\sigma} c^{\phantom{\dagger}}_{i,2,\sigma}+{\mathrm H.c.}\right)\nonumber \\
& + & U\sum_{i,\lambda}n_{i,\lambda,\uparrow} n_{i,\lambda,\downarrow}.
\end{eqnarray}
In this expression, $c^\dagger_{i,\lambda\sigma}$ ($c^{\phantom{\dagger}}_{i+1,\lambda\sigma}$) create (anihilate) a fermion with spin $\sigma$ on leg $\lambda$, and $U$ quantifies the on-site Coulomb interaction, which we take negative (attractive). In the rest of this work we consider $t_{x\uparrow}=t_{y\downarrow}=1$, $t_{x\downarrow}=t_{y\uparrow}=\alpha$, defining all energies in units of the hopping $t_{x\uparrow}$. In the case of two-leg ladders, the leg index assumes the values $\lambda=1,2$.   We also define the total fermion density as $n = \sum_\sigma \langle n_{i,\lambda \sigma} \rangle$, which lies in the range zero to two.  Due to a particle-hole symmetry
which takes $n \rightarrow 2 - n$, without loss of generality we can, and will, take 
$n$ between zero and one.

\begin{figure}
\epsfig {file=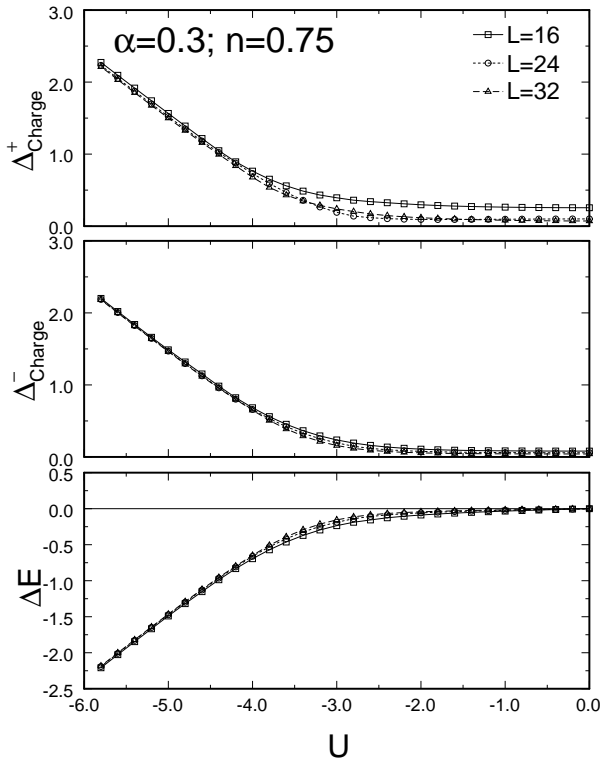,width=80mm}
\caption{
Charge gaps for $\alpha=0.3$ and $n=0.75$ as defined in the text, as a function of the attraction $U$ and for different system sizes. We also show the binding energy in the lower panel.
} \label{gap_charge_vs_U}
\end{figure}

\begin{figure}
\epsfig {file=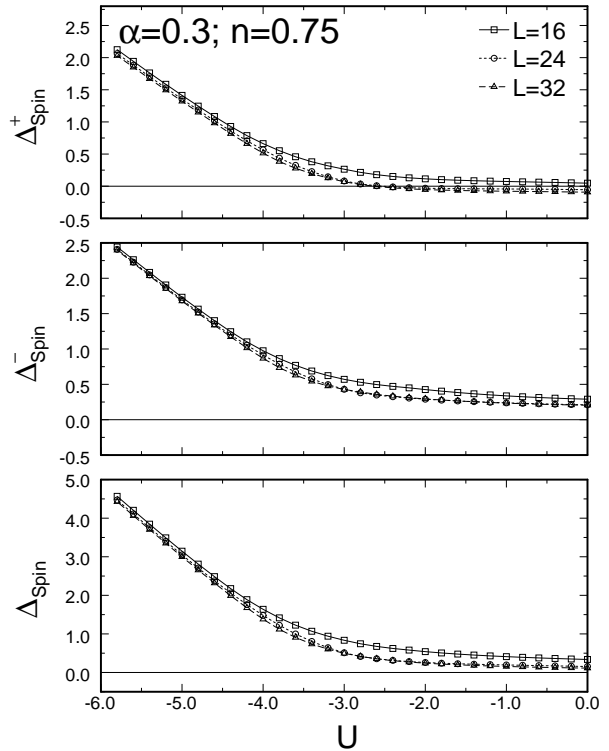,width=80mm}
\caption{
Spin gaps as defined in the text, for $\alpha=0.3$ and $n=0.75$, as a function of the attraction $U$ and for different system sizes.
} \label{gap_spin_vs_U}
\end{figure}

As a reference we consider the non-interacting limit ($U=0$) of this model. 
We shall pick a convention to denote what bands are partially filled, or depleted, as a function of the anisotropy and filling fraction, 
The different possibilities for an unpolarized mixture are depicted in Fig.\ref{bands}. The three ``phases'' are labeled by the number of bands that are partially filled, for each orientation of the spin, $(m_\uparrow,m_\downarrow)$, where $m_\sigma$ can assume the values $1$, or $2$. For instance, $(1,1)$ means that the bonding bands for both the $\uparrow$ and $\downarrow$ species are partially filled, while the anti-bonding bands are empty. It is important to notice that the Hamiltonian without interactions and finite anisotopy $0 < \alpha < 1$ has a ground state with finite polarization, as shown in Fig.\ref{bands}. This is easy to understand, and it is essentially due to peculiar band structure arising from the geometry we have considered: A majority of $\uparrow$-fermions would gain kinetic energy, since they have larger hopping along the leg direction. Therefore, it is natural to expect a ground state with $S^z_{Tot} > 0$. In some regimes the polarization is negative: for large anisotropy (small $\alpha$), the band for spin-$\downarrow$ is very flat, and it fills up very quickly as we increase the number of particles. However, 
we will primarily be interested in strong enough attractive interaction to pair all of the fermions
into a state with zero polarization.

As customary in most DMRG calculations, we take open boundary conditions along the leg direction, which improves convergence, and reduces calculation time.

\section{Results}

 We are interested in establishing and characterizing the various phases
 which appear in the model.  The parameters in the model are the hopping anisotropy, $\alpha$,
 the filling factor $n$ and the Hubbard attractive $U$ measured in units of the hopping strength, $t$.
 Since our main goal is to access the Cooper-pair Bose-metal phase, we will focus primarily
 on the regions of the $U=0$ ``phase diagram" in Figure \ref{bands} labelled $(2,1)$ wherein
 the up fermion has two partially filled bands and the down fermion only one.
 This corresponds to a regime of ``extreme" Fermi surface anisotropy.  More specifically,
 we will often report results for $\alpha = 0.3$ and $n=0.75$.
 We will then be interested in the accessible phases as the Hubbard $U$ is systematically increased.

 In order to characterize the different phases, we shall define several quantities of interest.
 We define the charge gap as the sum of the energies required to extract and inject a fermion into the system.
Since our model breaks $SU(2)$ symmetry, we can define
\begin{eqnarray}
\Delta^+_c & = & E_{(N+1,S+\frac{1}{2})}+E_{(N-1,S-\frac{1}{2})}-2E_{(N,S)}, \nonumber \\
\Delta^-_c & = & E_{(N+1,S-\frac{1}{2})}+E_{(N-1,S+\frac{1}{2})}-2E_{(N,S)}. \nonumber \\
\end{eqnarray}
 The spin gap is defined as the energy required to flip a spin. Similarly we can have
\begin{eqnarray}
\Delta^+_s & = & E_{(N,S+1)} - E_{(N,S)}, \nonumber \\
\Delta^-_s & = & E_{(N,S-1)} - E_{(N,S)} , \\
\Delta_s & = & \Delta^+_s + \Delta^-_s \nonumber .\\
\end{eqnarray}

\begin{figure}
\epsfig {file=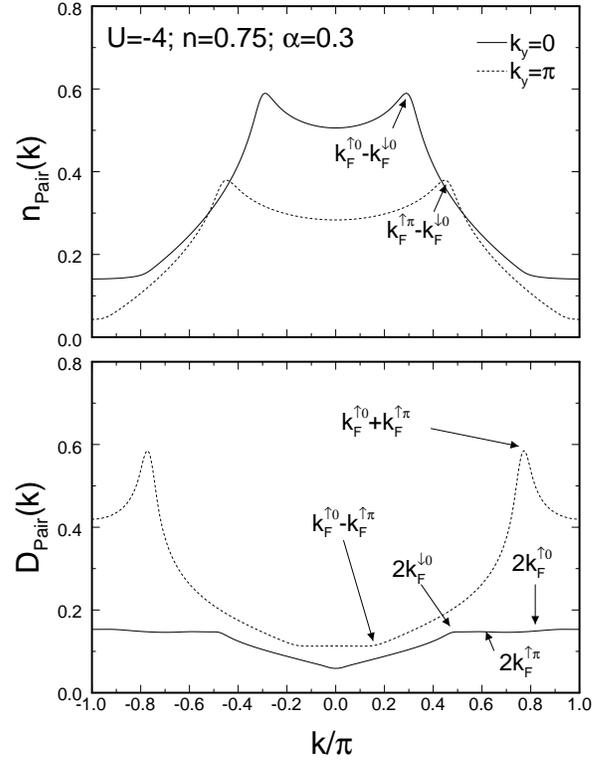,width=80mm}
\caption{Pair momentum distribution function and pair density structure factors for $L=48$ and parameters corresponding to the CPBM phase} 
\label{n_0.75.ty_0.3.U-4}
\end{figure}

We can also define the binding energy as the energy required to break a pair
\begin{eqnarray}
\Delta E & = & [E_{(N-2,S)}-E_{(N,S)}]-[E_{(N-1,S+\frac{1}{2})}-E_{(N,S)}] \nonumber \\
& - & [E_{(N-1,S-\frac{1}{2})}-E_{(N,S)}] \\
& = & E_{(N-2,S)}+E_{(N,S)}-E_{(N-1,S+\frac{1}{2})}-E_{(N-1,S-\frac{1}{2})}. \nonumber
\end{eqnarray}
Here, the first difference corresponds to the energy required to remove a pair, and the second (third) differences, the energy required to remove a single spin up (down) fermion. If the particles minimize their energy by creating a bound state, $\Delta E$ is negative, whereas for two independent particles $\Delta E = 0$ in the thermodynamic limit. In the case where the particles repel each other, this quantity is positive.

In Figure \ref{gap_charge_vs_U} we show results for the charge gap as a function of the attraction $U$.   We show results for the binding energy in the lower panel of Fig.\ref{gap_charge_vs_U}.
Figure \ref{gap_spin_vs_U} show results for the spin gaps, as a function of the attraction $U$. 

\subsection{``Metallic" state}

We first focus on the values of $|U| < 3$.  Here the charge gap vanishes in the thermodynamic limit
indicative of gapless fermion excitations.  Moreover, the binding energy, $\Delta E$ is very close
to zero suggesting an unpaired phase.  Finally, the spin gap also appears to vanish for these values
of $U$ in the large system size limit.   The gap $\Delta^+$ seems to show a tendency towards negative values, indicating that this phase may in fact have a small polarization. 
These results strongly suggest that for $|U| < 3$ the system is in a ``metallic" phase that is smoothly connected to the $U=0$ state, except with Luttinger liquid exponents characterizing the three gapless modes.

\begin{figure}
\epsfig {file=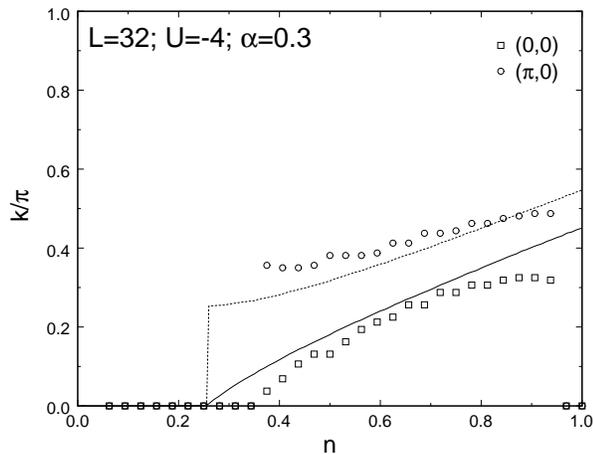,height=80mm,angle=-90}
\caption{
Position of the singular momenta $Q_x$ in the CPBM phase as a function of the density, for fixed anisotropy $\alpha=0.3$ and $U=-4$. Results are for a system with $L=32$, the same used for determining the phase diagram. Lines correspond to the prediction from the non-interacting picture, while symbols are DMRG results. The two lines corresponds to pairing between fermions in different bands, ($k_F^{\uparrow 0} - k_F^{\downarrow 0}$ and $k_F^{\uparrow \pi} - k_F^{\downarrow 0}$, respectively).
.
} \label{kmax.U-4}
\end{figure}

\subsection{Paired states}

For larger strengths of the attractive interaction, $|U| > 3$, there is a tendency for charge and spin gaps to open.  Moreover, the binding energy becomes negative indicating that all of the fermions
are bound into Cooper pairs.   It is natural to guess that once the fermions pair that they will
``condense" into a quasi-1D superfluid phase.  But as we now demonstrate, for intermediate
values of $U$ this appears not to be the case.

To characterize the nature of the paired state it is convenient to consider various correlation functions.
These are conveniently constructed from the onsite Cooper pair 
creation and annihilation operators
 $b^\dagger_i=c^\dagger_{i\uparrow}c^\dagger_{i\downarrow}$ and $b_i=c_{i\downarrow}c_{i\uparrow}$.
In the low-density limit the pairs behave in good approximation like canonical bosons
since $[b_i,b^\dagger_i] = 1-n_i \approx 1$. In other regimes, these will not be canonical bosons, but will give an indication of the nature of the Cooper pair excitations in the system.

Following this observation we define
the pair momentum distribution function (PMDF)
\begin{equation}
n_{\mathrm Pair}({\bf k}) = (1/L)\sum_{ij} \mbox{exp}[i{\bf k}({\bf r}_i-{\bf r}_j)]\, \langle b^{\dagger}_i b_j \rangle,
\label{nkpair}
\end{equation}
and the density structure factor
\begin{equation}
D_{\mathrm{Pair}}({\bf k}) = (1/L)\sum_{ij} \mbox{exp}[i{\bf k}({\bf r}_i-{\bf r}_j)]\, \langle n_{bi} n_{bj} \rangle.
\label{Dpair}
\end{equation}
Here, the Cooper pair number operator is defined as 
$n_{bi} = b^{\dagger}_i b_i = n_{i \uparrow} n_{i \downarrow}$.

\begin{figure}
\epsfig {file=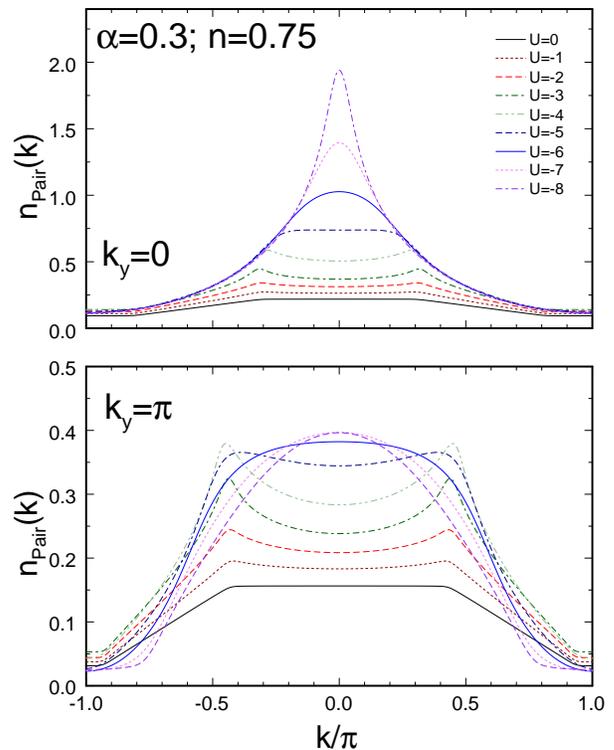,width=80mm}
\caption{
Pair momentum distribution for $\alpha=0.3$; $n=0.75$, and different values of $U$, showing the transition from metal to CPBM-like to superfluid at large values of $U$. The upper panel corresponds to momentum $k_y=0$, and the lower panel to $k_y=\pi$. Results are for a ladder of length $L=48$.
} \label{pair_vs_u}
\end{figure}

\begin{figure}
\epsfig {file=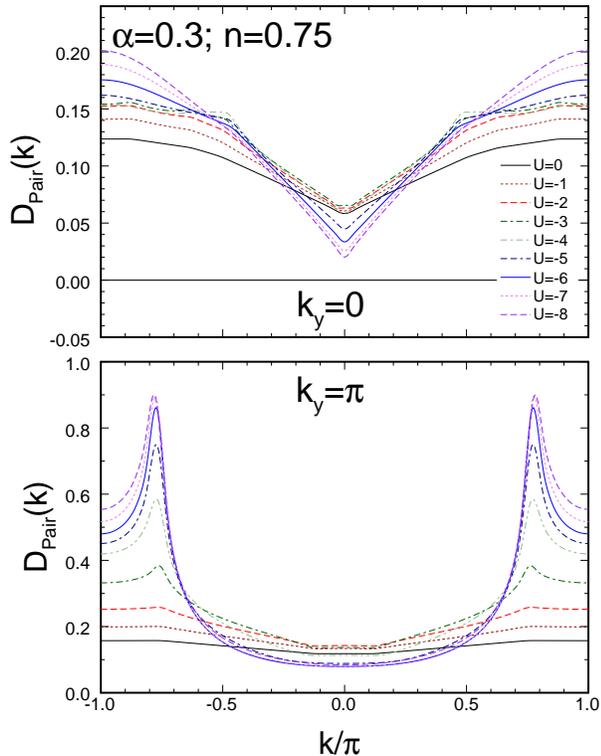,width=80mm}
\caption{
Pair density structure factor $\alpha=0.3$; $n=0.75$, and different values of $U$, showing the transition from metal to CPBM-like to superfluid at large values of $U$.
} \label{double_vs_u}
\end{figure}

\subsubsection{Cooper pair Bose-Metal}

We now focus on $|U|=4$ where the fermions are presumably only ``weakly" bound into Cooper pairs.
In Fig.\ref{n_0.75.ty_0.3.U-4} we show the pair momentum distribution function and the pair density structure factors.   Most striking are the finite momentum singular features in both correlators.
We can proceed to analyze all the singularities in both the PMDF and pair structure factor following \cite{Sheng08}.  
In both cases, we can trace the position of the peaks by just looking at the prediction from the non-interacting band structure.

Consider first the pair momentum distribution function.  Due to the mismatched Fermi surfaces,
up fermions and down fermions cannot pair at zero momentum.
Rather, there will be pairing tendencies at 
finite center of mass momentum, ${\bf Q}={\bf k}_{F}^{\uparrow}-{\bf k}_{F}^{\downarrow}$, where the Fermi momenta are defined with respect to the non-interacting dispersion. 
Specifically, a pair with zero $y-$component center of mass, $Q_y=0$, can be formed
by combining a right moving up spin fermion from the bonding band, with $x-$momentum,
$k_F^{\uparrow 0}$, with a left moving spin down fermion from the bonding band
with momentum $-k_F^{\downarrow 0}$.  As shown on the top panel of Figure 4,
the resultant center of mass momentum $Q_x = k_F^{\uparrow 0} - k_F^{\downarrow 0}$
corresponds nicely to the location of the peak.  Similarly, the peak at center of mass momentum
$Q_y = \pi$ results from a right moving up spin fermion from the anti-bonding band,
$k_F^{\uparrow \pi}$, ``pairing" with a left moving down spin from the bonding band.    

It must be emphasized that these singular features that ``know" about the non-interacting Fermi surface are appearing in the pair correlator despite the fact that all of the fermions are bound into Cooper pairs and the system has a fermionic (charge) gap!   This most surprising feature
is a hallmark of the Cooper-pair Bose metal.  Propagating Cooper pairs moving thru a fluid with a fermion gap are somehow still sensitive to the underlying Fermi surfaces of the constituent 
particles.   

In Figure \ref{kmax.U-4} we show the comparison with this theoretical prediction for a fixed value of $U=-4$ and $\alpha=0.3$, as a function of the density $n$. As the band fillings change, the Cooper pair momenta change accordingly. Again, deep in the CPBM phase, the pairs keep memory of the non-interacting Fermi surfaces.

In the lower panel of Fig.\ref{n_0.75.ty_0.3.U-4} we show the pair density structure factors.
Again, there are a number of singular features in momentum space.   Since the
fermions are all paired in this regime, one might anticipate that the density fluctuations
of the  up and down spin fermions are identical to one another, and equal to the pair density
fluctuations.  But this is not the case.  Rather, density fluctuations of the fermions differ,
and both appear to contribute to the pair density structure factor.  In the figure we have demarcated
various $``2k_F"$ momenta constructed from the non-interacting dispersion of the fermions.
Some of the features line up with the singularities remarkably well.

\begin{figure}
\epsfig {file=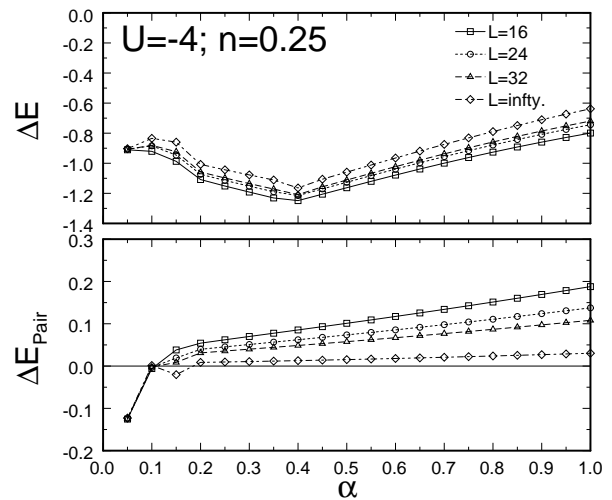,width=80mm}
\caption{
Finite size study of the binding energy for $U=-4$ and $n=0.25$. We show results for $L=16,24,32$ and a extrapolation to $L=\infty$. From top to bottom we show the binding energy for fermions, and the binding energy for pairs.
} \label{binding_U-4_n0.25}
\end{figure}

\begin{figure}
\epsfig {file=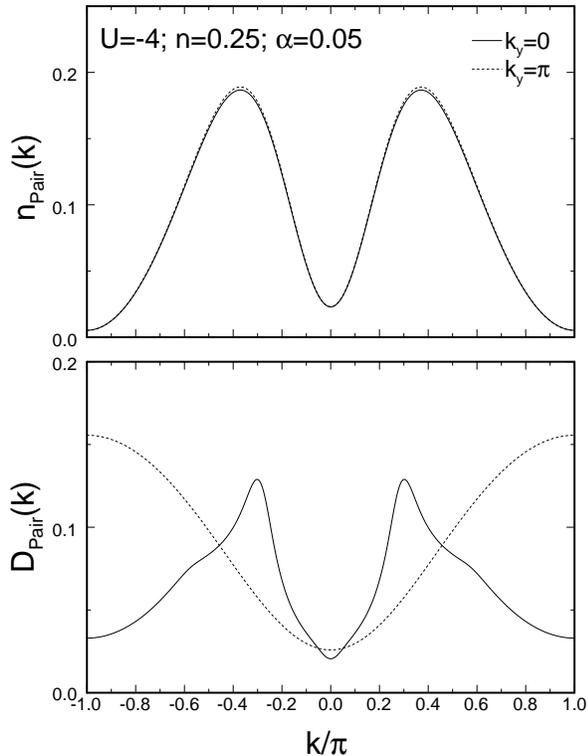,width=80mm}
\caption{Pair momentum distribution function and pair density structure factors for $L=48$ and parameters corresponding to d-wave bosonic pairs.} 
\label{n_0.25.ty_0.05.U-4}
\end{figure}

\subsubsection{Superfluid} 
 
 To show the evolution to the conventional superfluid out of the Cooper-pair Bose metal phase,
 we can continue to increase $|U|$.   
 Figure \ref{pair_vs_u} shows the PMDF for density $n=0.75$, $\alpha=0.3$ for a system  of length $L=48$ at various different values of $U$.
 As $U$ increases from zero, the non-interacting curves start developing
a two-peaked structure both at $k_y=0$ and $k_y=\pi$, with singularities at finite momentum. 
The two-peaked structure is maximum near $U=-4$, where the system is in the Cooper-pair
Bose metal phase.   But for still larger $U$ somewhere in the region $5 < |U| < 6$,  the two-peaked structure gradually evolves
into a single peak at zero momentum.  

At the largest value $U=-8$ we are presumably in the conventional quasi-1d superfluid.
Indeed, the most prominent feature is a large peak at momentum ${\bf Q} = (0,0)$,
which continues to grow with increasing $U$.  This is 
indicative of a quasi-condensate.  The pair momentum distribution function
at $k_y=\pi$, on the other hand, appears to have saturated with increasing $U$
and shows a rather smooth structure throughout the momentum space.   For bosons moving on a 2-leg ladder these are the expected signatures of a quasi-1d superfluid (see Fig. 5 in Ref.\onlinecite{Sheng08}).

A similar change of behavior is observed in the singularities of the pair density structure factor, Fig.\ref{double_vs_u}. For small $|U|$, the $k_y=0$ component shows a linear behavior near $k_x=0$ and kinks or singularities at finite momentum, which are largest near $U=-4$ in the Cooper pair Bose-metal phase (and also    
observed in the DBM phase  - compare to Fig. 8 in Ref.\onlinecite{Sheng08}). For the largest values of $U$ the density structure factor at $k_y=0$ has a v-shape, being quite smooth away from zero momentum,
as expected in a superfluid.  

At $k_y=\pi$ the singular features at $U=-4$ in the Cooper-pair Bose-metal evolve into rather large peaks with increasing $U$.  The peak height appears
to saturate at the largest value of $U$, and is perhaps becoming smoother.
The behavior here is somewhat puzzling, since a quasi-1d superfluid should have a pair
distribution function at $k_y=\pi$ which is analytic in $k_x$ (see, for example, Fig. 5 in Ref.\onlinecite{Sheng08}).
Ideally, one would try to obtain data for increasing system size to see if the behavior in the superfluid
regime saturates and smoothens.  One would expect the data at $U=-4$ in the CPBM to become more singular in this limit.

\subsubsection{Pairing of Cooper pairs?}

We further explored a wider region of parameter space, both varying density and hopping anisotropy. Building on predictions from \cite{Motrunich07,Sheng08}, we have anticipated the possibility of a phase of paired Cooper pairs, or paired bosons. The bosonic ring models indeed display such phases, and bosons can pair with both s-wave symmetry, or d-wave symmetry. A state with paired Cooper pairs would in turn have a finite binding energy {\it for breaking a pair of Cooper pairs}.
 We can define the binding energy for bosonic pairs (Cooper pairs) as
\begin{eqnarray}
\Delta E_{Pair} = E_{(N-4,S)}+E_{(N,S)}-2E_{(N-2,S)}.
\end{eqnarray}

Fig. \ref{binding_U-4_n0.25} shows our results for the pair binding energy as a function of anisotropy $\alpha$ for a fixed value of interaction $U=-4$ and density $n=0.25$. We find a negative binding energy for the fermions in all the range of $\alpha$, clear indication of pairing. The bosonic binding energy is indeed positive or very small for large $\alpha > 0.1$, but for large anisotropy (small $\alpha$), it dramatically turns negative. This seem to indicate the presence of a new exotic phase with paired Cooper pairs.

In order to characterize this phase we have looked at the PMDF as well as the pair density structure factor. Fig.\ref{n_0.25.ty_0.05.U-4} shows our results for $\alpha=0.05$ and $L=48$. By comparing with the prediction from Ref.\onlinecite{Sheng08}, we conclude that this profile corresponds to the d-wave paired state of Cooper pairs.
(Compare to their Fig.14.)   Specifically, since the Cooper pairs are paired, one expects a gap for the single Cooper pairs.   This corresponds to a smooth pair momentum distribution function, as indeed
seen in Figure \ref{n_0.25.ty_0.05.U-4}.   Moreover, at density $n=0.25$, the distance between pairs of Cooper pairs down the ladder is $2/n = 8$ sites.  One would then expect that
the density structure factor at $q_y=0$ would show a singular feature at wavevector
$2 \pi/8 = \pi/4$.  This singular feature is indeed prominent in Figure \ref{n_0.25.ty_0.05.U-4}.  

\begin{figure}
\epsfig {file=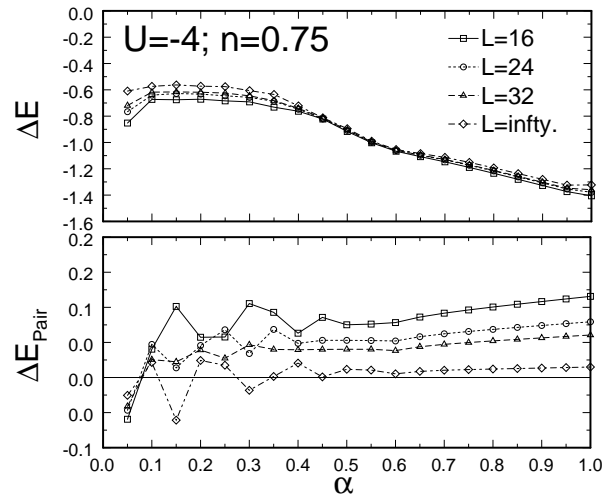,width=80mm}
\caption{
Finite size study of the binding energy for $U=-4$ and $n=0.75$. We show results for $L=16,24,32$ and a extrapolation to $L=\infty$. From top to bottom we show the binding energy for fermions, and the binding energy for pairs.
Again, notice the size-dependent oscillatory behavior in the CPBM region.
} \label{binding_U-4_n0.75}
\end{figure}

We have done a similar analysis of binding energies for $n=0.75$, shown in Fig.\ref{binding_U-4_n0.75}. For values of anisotropy $\alpha > 0.5$, the results correspond to a superfluid phase. At intermediate values $0.1 < \alpha < 0.5$ we found a wildly oscillatory behavior in the bosonic binding energy, accompanied by strong finite size effects. We attribute this behavior to the CPBM phase. The Cooper pair structure in this phase is strongly dependent on the availability of momenta to pair, which varies in finite systems for different system sizes. At small values of $\alpha$ we found again the d-wave bosonic paired state.

\section{bosonization}

\begin{figure}
\epsfig {file=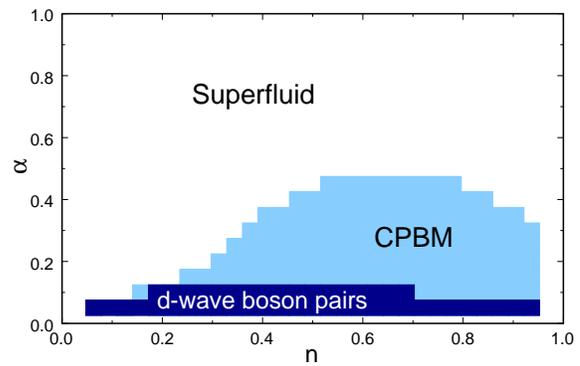,height=80mm,angle=-90}
\caption{
Phase diagram of the Hubbard ladder with anisotropic hopping $\alpha$, and length $L=32$, $U=-4$, as a function of the density and anisotropy. We find a Cooper-pair Bose-metal phase (CPBM), and a superfluid phase. States of d-wave boson pairs are found at high density and small $\alpha$.
} \label{phase diagram U-4}
\end{figure}

This collection of results can be summarized in the phase diagram of Fig. \ref{phase diagram U-4}, as a function of density and anisotropy, for a fixed values of $U=-4$. The region of stability for the CPBM phase, roughly between $3 < |U| < 5$, shrinks with increasing $|U|$, leading to a conventional superfluid with pairing momentum ${\bf Q}=(0,0)$. 
At $U=-4$ it is striking that the region of the Cooper-pair Bose-metal phase roughly corresponds to the
region where the non-interacting band structure is in the $(2,1)$ regime.
Loosely, this can be understood via a bosonization analysis that we now briefly sketch.

We follow very closely
Ref. \onlinecite{Sheng08}, where a detailed analysis was performed by bosonizing the fermionic partons
which were introduced by decomposing the hard core boson as, $b = d_1 d_2$.
In that work the partons were coupled to a $U(1)$ gauge field that glued them back together.
Here we instead can bosonize directly the ``fundamental" fermions, $c_\sigma$
that enter into the Hamiltonian.   As we shall see, this will lead to the same description of the
DBM phase.

To proceed, when $U=0$ we can conveniently linearize the fermion bands about their
respective Fermi momentum, focussing on the slowly varying fields,
$c_{\sigma P}^{(k_y)}$, where $P=R/L= \pm$ corresponds to a right and left moving field, $\sigma = \uparrow,\downarrow$ the spin  and $k_y = 0,\pi$ labels the bonding/antibonding bands, respectively.  
We employ bosonization,
\begin{equation} 
c_{\sigma P}^{(k_y)} = \eta_\sigma^{(k_y)} \exp{[i(\phi_\sigma^{(k_y)} + P \theta_\sigma^{(k_y)})]},
\end{equation}
where $\phi,\partial_x \theta$ are conjugate fields and the $\eta's$ are Klein factors.
The non-interacting Lagrangian density can be expressed as,
\begin{equation}
{\cal L}_0 = \frac{1}{2\pi} \sum_{\sigma, k_y} [ v_{\sigma}^{(k_y)} ( \partial_x \theta_\sigma^{(k_y)} )^2 +
\frac{1}{v_\sigma^{(k_y)}} (\partial_\tau \theta_\sigma^{(k_y)} )^2]. 
\end{equation}

We now focus initially on the $(1,1)$
case where only the two bonding bands are partially filled.
In the case of zero polarization the Fermi wave vectors satisfy, 
$k_{F \uparrow}^{(0)} = k_{F \downarrow}^{(0)}$.
In the presence of an attractive $U$ there is then an allowed momentum conserving
four-fermion interaction in the Cooper channel,
\begin{equation}
{\cal H}_u = - u c_{R \uparrow}^{(0) \dagger } c_{L \downarrow}^{(0) \dagger} c_{R \downarrow }^{(0)} c_{L \uparrow}^{(0)} + H.c..
\end{equation}
This term can lead to a paired superfluid phase with a spin gap, as can be seen by bosonization, ${\cal H}_u \sim - u \cos[2(\theta_\uparrow^{(0)} - \theta_\downarrow^{(0)})]$.  Provided this term
is marginally relevant it grows under renormalization, and the cosine term can be expanded.
This gaps out the spin mode, $\theta_\uparrow - \theta_\downarrow$ and leads to a single gapless
mode which describes the quasi-1d superfluid state.  It is worth commenting that in the presence of a non-zero polarization in the $(1,1)$ regime, the Cooper channel is no longer momentum conserving.
Nevertheless, non-perturbatively one expects the attractive Hubbard $U$ to drive the system into
a superfluid, beyond some threshold.

Next consider the $(2,1)$ regime, focussing on the case with zero polarization, so that
the Fermi wavevectors satisfy, $k_{F \uparrow}^{(0)} + k_{F \uparrow}^{(\pi)}
= k_{F \downarrow}^{(0)}$.   In the presence of an attractive $U$, there are various allowed four-fermion interactions, but the Cooper channel is not
present due to a lack of nesting between the up and down spin Fermi wavevectors.  There is, however, an important momentum conserving $6-$fermion term of the form,
\begin{equation}
H_v = -v_6 c_{\uparrow R}^{(0) \dagger}  c_{\uparrow L}^{(0)} 
c_{\uparrow R}^{(\pi) \dagger} c_{\uparrow L}^{(\pi)} 
c_{\downarrow L}^{(0) \dagger}  c_{\downarrow R}^{(0)}  + H.c .
\end{equation}
Being 6th order this term will be irrelevant at weak coupling, 
and the system will be in a ``metallic" state with three gapless modes.
But at stronger coupling 
above a threshold value of $U$ when the forward scattering interactions shift the scaling dimension of $v_6$, this term can become relevant.
Under bosonization this term becomes, 
\begin{equation}
{\cal H}_v = -v_6 \cos[2(\theta_\uparrow^{(0)} + \theta_\uparrow^{(\pi)} - \theta_\downarrow^{(0)}) ] ,
\end{equation}
and above the threshold we can expand the cosine term to obtain a mass term for the combination,
$\theta_M = (\theta_\uparrow^{(0)} + \theta_\uparrow^{(\pi)} - \theta_\downarrow^{(0)})/\sqrt{3}$.
In the gauge theory analysis in Ref. \onlinecite{Sheng08}, just such a mass term is present due to the long-ranged
interactions mediated by the gauge field.  The main difference here is that the mass term is generated
via an instability driven, at intermediate coupling, by the attractive Hubbard $U$

Following Ref. \onlinecite{Sheng08} , upon integrating out $\theta_M$ one obtains a theory of two-coupled
Harmonic modes, $\theta_1 = (\theta_\uparrow^{(0)} + \theta_\uparrow^{(\pi)} +2 \theta_\downarrow^{(0)})/\sqrt{6}$ and $\theta_2 = (\theta_\uparrow^{(0)} - \theta_\uparrow^{(\pi)})/\sqrt{2}$.  
This is the fixed point description of the DBL $(2,1)$ phase,
what we are referring to as the Cooper-pair Bose metal.  

To evaluate correlators it is convenient to define 
new conjugate fields, $\phi_M, \phi_1,\phi_2$ in the same way.
Inverting the canonical transformation gives,
\begin{eqnarray}
\theta_\uparrow^{(0/\pi)} &=& \frac{1}{\sqrt{6}} \theta_1 \pm \frac{1}{\sqrt{2}} \theta_2 + \frac{1}{\sqrt{3}}\theta_M ,\\
\theta_\downarrow^{(0)} &=& \sqrt{\frac{2}{3}}  \theta_1 - \frac{1}{\sqrt{3}}\theta_M  ,
\end{eqnarray}
with identical expressions for the $\phi's$.
Using these 
one can readily show that
the bosonized expressions for the fermion operators always involve an exponential of $\phi_M$.
Since $\theta_M$ is massive, $\phi_M$ fluctuates wildly, and the fermion is gapped.  Moreover, the Cooper pair creation operators,
$c_{\uparrow P}^{(k_y)} c_{\downarrow P^\prime}^{(0)}$ are independent of $\phi_M$,
and will thus exhibit power law correlators.   These are properties of the Cooper-pair Bose metal.

\section{Summary and discussion}

In this paper we have explored the possible phases present in a model of fermions
hopping on a 2-leg ladder with spin dependent hopping strengths.  Our main conclusion
is the presence of an unusual Cooper-pair Bose-metal phase for intermediate values
of the attractive Hubbard $U$.   In this novel phase the fermions are fully gapped,
but the Cooper-pair operator is in a gapless state which is qualitatively distinct from
the quasi-1d superfluid \cite{Feiguin09b}. In particular, there are two
gapless modes in the Cooper-pair Bose-metal phase, in contrast to the conventional superfluid which has only one gapless mode.
Moreover, in the CPBM the Cooper pair momentum distribution function shows singularities both at $k_y$ equal
to zero and $\pi$ and at non-zero values of the longitudinal momentum, $k_x$.
By contrast, the quasi-1d superfluid has a Cooper pair momentum distribution function which is smooth at $k_y=\pi$,
and at $k_y=0$ has a singular peak at the origin, $k_x=0$.

In addition to the CPBM at intermediate $U$ and the superfluid phase at larger values of $U$,
the model studied here has a stable ``metallic" phase which is continuously connected to
the non-interacting (2,1) phase.   In this ``metallic" phase there are
three gapless modes, same as in the non-interacting limit.  The same occurrs 
in a quasi-1d LO state, which also has as many gapless modes as the reference non-interacting limit.
Analogs of both the CPBM and the superfluid are present
in the hard core boson hopping with ring exchange studied in Ref. \onlinecite{Sheng08}.
Here, the onsite Cooper pair operator is playing the role of the boson.
But the ``metallic" phase only can exist in the fermion model studied in this work.

To distinguish these three phases experimentally would require measuring both
the fermion momentum distribution function (singular in the metallic phase
and smooth in the CPBM and the superfluid) and the Cooper-pair momentum distribution function (singular in all three phases, but with a distinct signature in the superfluid).
The former could be measured by releasing the atoms from the trap in the usual way.
But to extract the Cooper-pair momentum distribution function would require a 
sudden quench to $|U| \rightarrow \infty$ just before releasing the atoms from the trap.

The low dimensionality of the ladder geometry studied here implies that the singularities in the pair momentum distribution function  can only appear at discrete points in momentum space. However, in two dimensions we expect \cite{Motrunich07} that the
Cooper-pair Bose-metal phase will exhibit a pair momentum distribution function that is
singular along  lines  or ``Bose surfaces'' in momentum space. In contrast, the only way to obtain a true Bose condensate in 2D is by a macroscopic condensate into a state with a single, or a finite discrete set of momenta ${\bf Q}$, such as the structures predicted for FFLO-like condensates \cite{Bowers02}. Due to the strong frustration in our model --responsible for the ring exchange term-- we are inclined to believe that the more likely scenario is the first one, with uncondensed Cooper pairs. However, an answer to this question would require an actual strong-coupling study of the two-dimensional system.

One of the most striking features of our study is the conclusion that fermions with spin-dependent anisotropic Fermi surfaces and attractive interactions behave very much like hard-core bosons with a ring exchange, giving rise to much of the same physics already observed in these models \cite{Motrunich07,Sheng08}. In the case of hard-core bosons, the wave function is accurately described by fractionalizing the bosons into two partons, or fermions with anisotropic fermi surfaces.
However, these partons are a fictitious construction while our fermions are real. This implies that
``constructing" frustrated boson systems with , for example, ring exchange interactions
might be much easier by pairing underlying spinful fermions than by working directly with bosonic atoms.   Indeed, as proposed in Ref. \onlinecite{Feiguin09}, a cold Fermi gas of Yb atoms \cite{Fukuhara07} loaded in a spin-dependent optical lattice subject to an attractive s-wave potential\cite{Mandel03,Mandel03b,Liu04} might work just as well as the proposed setup for generating ring exchange interactions in a system of hard-core bosons \cite{Buchler05}. 

An alternative setup to realize exotic paired states could be achieved by using a Fermi mixture with different atomic species, such as 
 Li and K for instance, with inter-species Feshbach tunable interactions.
This would give one the ability to make spin dependent optical lattices  
since the two atoms are distinct and so are much easier to
independently optically control than merely distinct hyperfine states  
of the same atom (which tend to have similar polarizability
and so see a much more similar optical potential). Moreover, one may only use an optical lattice for  only one of the species, leaving the other basically free.

If --contrary to our argument-- a 2D system of fermions with spin-dependent hopping indeed undergoes a true Bose condensation, the condensate could be described by a nodal structure that would have similar characteristics as the one for ``striped superconductivity'' or pair-density wave
order. This type of order, proposed in Refs.\onlinecite{Berg08,Berg09} to account for experimental observations in La$_{2-x}$Ba$_x$CuO$_4$ \cite{Li07}, would actually break rotational symmetry. In Ref.\onlinecite{Berg09b} the authors argue that a thermal melting of the stripe superconducting state could give rise to $4e$ superconductivity originating from the coupling between condensates with perpendicular stripe order, similar to the $4e$ superconductivity we see in the very anisotropic regime of our ladder model. We point out that a similar behavior has been predicted to occur in polarized mixtures \cite{Leo and Ashvin}. At finite temperatures the LO state is always unstable to a nematic superfluid. Fluctuations can destroy the superfluid leading to a state of paired Cooper pairs.

Finally, we want to point out that we have not considered the possibility of phase separation in the present study\cite{Leo,Leo2}. Even though it may in fact occur, as it happens in the BCS mean field treatment of the model in 2D \cite{Feiguin09}, we suspect that it will only take place in narrow regions of the phase diagram separating the different phases, and will not be a dominating feature.

\section{Acknowledgements}

We would like to thank Lesik Motrunich for numerous fruitful discussions
about the present study, and for earlier collaborations leading up to this work.
Thanks are also due to Victor Gurarie, Leo Radzihovsky, and Peter Zoller for helpful discussions. 
The authors are grateful to the NSF for support under grants  No. DMR-0529399 (MPAF), and DMR-0955707 (AEF).

\end{document}